\documentclass[amsmath,pra,twocolumn,showpacs]{revtex4}
\usepackage{graphics}

\begin{document}     

\title{Spontaneous radiative decay of translational levels 
of an atom near a dielectric surface}

\author{Fam Le Kien} 
\altaffiliation{Also at Institute of Physics and Electronics, Vietnamese Academy of Science and Technology, Hanoi, Vietnam.}
\affiliation{Department of Applied Physics and Chemistry, University of Electro-Communications, Chofu, Tokyo 182-8585, Japan}
\author{K. Hakuta} 
\affiliation{Department of Applied Physics and Chemistry, University of Electro-Communications, Chofu, Tokyo 182-8585, Japan}

\date{\today}

\begin{abstract}
We study spontaneous radiative decay of translational levels of an atom in the vicinity of a semi-infinite dielectric. We systematically derive the microscopic dynamical equations for the spontaneous decay process. We calculate analytically and numerically the radiative linewidths and the spontaneous transition rates for the translational levels.
The roles of the interference between the emitted and reflected fields 
and of the transmission into the evanescent modes are clearly identified. 
Our numerical calculations for the silica--cesium interaction show that the radiative linewidths of the bound excited levels with large enough but not too large vibrational quantum numbers are moderately enhanced by the emission into the evanescent modes and 
those for the deep bound levels are substantially reduced by the surface-induced red shift of the transition frequency.
\end{abstract}

\maketitle

\section{Introduction}
\label{sec:introduction}

Due to recent progress in atom optics and nanotechnology, the study of individual neutral atoms in the vicinities of material surfaces has gained renewed interest 
\cite{our traps,Lima,Oria2006,cesium decay,absorption}.
A method for microscopic trapping and guiding individual atoms along a nanofiber has been proposed \cite{our traps}. An optical technique for loading atoms into
quantum adsorption states of a dielectric surface has been suggested \cite{Lima,Oria2006}.
The possibility to control and manipulate individual atoms near surfaces can find applications for quantum information \cite{Schlosser,Kuhr,Sackett} and  atom chips \cite{Folman,Eriksson}. Cold atoms can be used as a probe that is very sensitive to the surface-induced perturbations
\cite{surface probe}. 
Such state-of-the-art applications require the development of a systematic microscopic theory 
for spontaneous radiative decay of atoms near surfaces.

It is well known that the spontaneous emission rate of an atom is modified by the presence of an interface \cite{response,Agarwal,Wylie,conductor,dielectric}. 
A quantum-mechanical linear-response formalism has been
developed for an atom close to an arbitrary interface \cite{response,Agarwal,Wylie}.  
An alternative approach based on the mode expansion has been used for an atom near a perfect conductor \cite{conductor} or a dielectric \cite{dielectric}. 
In the previous treatments \cite{response,Agarwal,Wylie,conductor,dielectric},
the effect of the center-of-mass motion of the atom
on the spontaneous emission process was neglected. In this condition, the effects
of the surface on the spontaneous radiative decay of the atom manifest simply as a quantum 
electrodynamic enhancement due to the mode modification
and a frequency shift due to the surface--atom van der Waals interaction,
with no structural changes in the dynamics of the process  \cite{Agarwal,Wylie}.

Lately translational levels of an atom in a surface-induced potential have been 
studied \cite{Lima,Oria2006}.
An experimental observation of the excitation spectrum of cesium atoms in quantum adsorption states of a nanofiber surface has been reported \cite{Kali}.
In the theoretical treatments of Refs. \cite{Lima,Oria2006}, the quantum electrodynamic
enhancement of spontaneous emission due to the mode modification was completely neglected,
and a set of equations with phenomenologically added decay terms was used for
describing the dynamics of the loading process. A more rigorous theory requires
a deeper knowledge of the structure of the decay equations and the magnitudes of the decay coefficients for the translational levels of the atom.

In this paper, we study spontaneous radiative decay of translational levels of an atom in the vicinity of a semi-infinite dielectric. Our study is general in the sense that it incorporates the quantum nature of the radiation field in the presence of the dielectric as well as the quantum center-of-mass motion of the atom. 
We systematically derive the microscopic dynamical equations for the spontaneous decay process, which reflect the complexity of the translational levels of the atom in the potential. We focus our attention on the radiative linewidths and the spontaneous transition rates. Our results obtained by the quantum mode-expansion approach allow us to identify
the roles of the interference between the emitted and reflected fields 
and of the transmission into the evanescent modes. 

The paper is organized as follows. In Sec.\ \ref{sec:model} we describe the model.
In Sec.\ \ref{sec:analytical} we derive the basic dynamical equations for the spontaneous radiative decay process. 
In Sec.\ \ref{sec:rates} we study analytically the decay rates.
In Sec.\ \ref{sec:numerical} we present the results of numerical calculations. 
Our conclusions are given in Sec.~\ref{sec:summary}.

\section{Model}
\label{sec:model}

We consider a space with one interface. 
The half-space $x<0$ is occupied by a nondispersive nonabsorbing dielectric medium (medium 1).
The half-space $x>0$ is occupied by vacuum (medium 2).
We examine an atom, with an upper internal level $e$ and a lower internal level $g$, moving in the empty half-space $x>0$. 

\subsection{Quantum translational states of the atom}

\begin{figure}[tbh]
\begin{center}
  \includegraphics{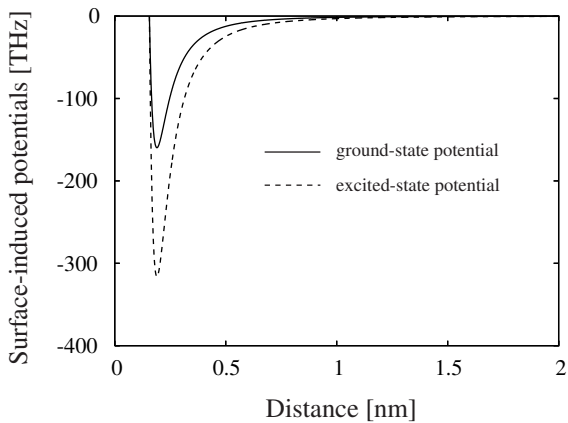}
 \end{center}
\caption{Surface-induced potentials for ground- and excited-state atoms. 
The parameters are
$C_{3g}=1.56$ kHz $\mu$m$^3$, $C_{3e}=3.09$ kHz $\mu$m$^3$, 
$A_g=1.6\times 10^{18}$ Hz, $A_e=3.17\times 10^{18}$ Hz, and $\alpha_g=\alpha_e=53$ nm$^{-1}$.
They are taken for the case of a $D_2$-line cesium atom in the vicinity of a semi-infinite silica medium.
}
\label{fig1}
\end{figure}

We describe the quantum translational (center-of-mass) states of the atom by
using the formalism of Ref. \cite{Lima}.
The potential energy of the surface--atom interaction is
a combination of a long-range van der Waals attraction $-C_3/x^3$ and
a short-range repulsion \cite{Hoinkes}. 
Here, $C_3$ is the van der Waals coefficient.
We approximate the short-range repulsion by an exponential function $Ae^{-\alpha x}$,
where the parameters $A$ and $\alpha$ determine the height and range, respectively, of the repulsion.
The combined potential depends on the internal state of the atom (see Fig. \ref{fig1}), and
is presented in the form
\begin{equation}
V_j(x)=A_j e^{-\alpha_j x}-\frac{C_{3j}}{x^3},
\label{24}
\end{equation} 
where $j=e$ or $g$ labels the internal state of the atom.

The potential parameters $C_{3j}$, $A_j$, and $\alpha_j$ depend 
on the dielectric and the atom. 
In numerical calculations, we use the parameters of fused silica, for the dielectric,
and the parameters of atomic cesium with the $D_2$ line, for the two-level atomic model.
In Appendix \ref{sec:potential}, 
the parameters of the ground- and excited-state potentials for the silica--cesium interaction are estimated to be 
$C_{3g}=1.56$ kHz $\mu$m$^3$, $C_{3e}=3.09$ kHz $\mu$m$^3$, 
$A_g=1.6\times 10^{18}$ Hz, $A_e=3.17\times 10^{18}$ Hz, 
and $\alpha_g=\alpha_e=53$ nm$^{-1}$.  
The potential depths are $D_g=159.6$ THz and $D_e=316$ THz. 
The minimum positions for both potentials 
are assumed to be the same, equal to $x_m=0.19$ nm. 

The Hamiltonian of the atom in the surface-induced potential is given by
\begin{equation}
H_A=\frac{p^2}{2m}+\sum_{j=e,g}[\hbar\omega_j+V_j(x)]|j\rangle\langle j|.
\label{1}
\end{equation}
Here, $p$ and $m$ are the momentum and mass of the atom, respectively,
and $\omega_j$ is the frequency of the internal level $j$. 
We introduce the notation $|\varphi_{a}\rangle$ and $|\varphi_{b}\rangle$  
for the  eigenstates of the center-of-mass motion of the atom in the potentials
$V_e(x)$ and $V_g(x)$, respectively.  
The wavefunctions $\varphi_{a}(x)$ and $\varphi_{b}(x)$ 
are determined by the stationary Schr\"{o}dinger equations
\begin{eqnarray}
\left[-\frac{\hbar^2}{2m}\frac{d^2}{dx^2}+V_e(x)\right]\varphi_{a}(x)
&=&\mathcal{E}_{a}\varphi_{a}(x),
\nonumber\\
\left[-\frac{\hbar^2}{2m}\frac{d^2}{dx^2}+V_g(x)\right]\varphi_{b}(x)
&=&\mathcal{E}_{b}\varphi_{b}(x),
\label{2}
\end{eqnarray}
respectively. The eigenvalues $\mathcal{E}_{a}$ and $\mathcal{E}_{b}$
characterize the energies of the translational levels of the excited and ground states, respectively. Without the loss of generality, we assume that
the center-of-mass eigenfunctions $\varphi_{a}$ and $\varphi_{b}$ are real functions.

\begin{figure}[tbh]
\begin{center}
 \end{center}
\caption{Energies and wave functions of the center-of-mass motion of the ground- and excited-state atoms in the surface-induced potentials. 
The parameters of the potentials are as in Fig. \ref{fig1}. 
The mass of  atomic cesium $m=132.9$ a.u. is used.
We plot only the bound levels of the ground and excited states with energies 
in the range from $-1$ GHz to $-5$ MHz and the free ground-state level with energy of about 4.25 MHz.
}
\label{fig2}
\end{figure} 

\begin{figure}[tbh]
\begin{center}
  \includegraphics{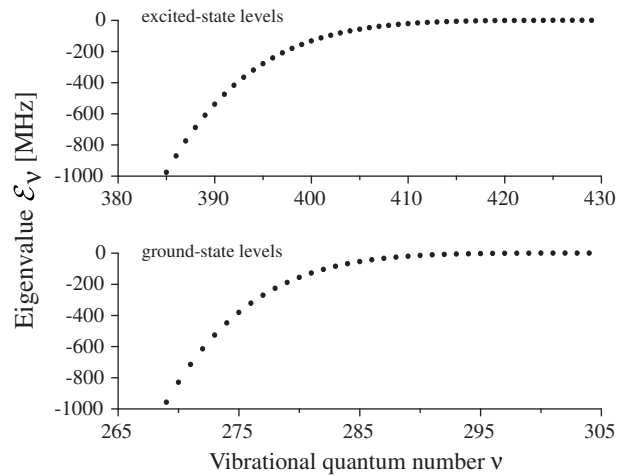}
 \end{center}
\caption{Eigenvalues for the bound levels of the ground and excited states of the atom in the range from $-1$ GHz to $-20$ kHz. 
The parameters used are as for Fig. \ref{fig2}.
}
\label{fig3}
\end{figure}

\begin{figure}[tbh]
\begin{center}
 \end{center}
\caption{Same as Fig. \ref{fig2} but for the twenty lowest levels
of the excited state and the forty lowest levels of the ground state.
}
\label{fig4}
\end{figure} 

\begin{figure}[tbh]
\begin{center}
  \includegraphics{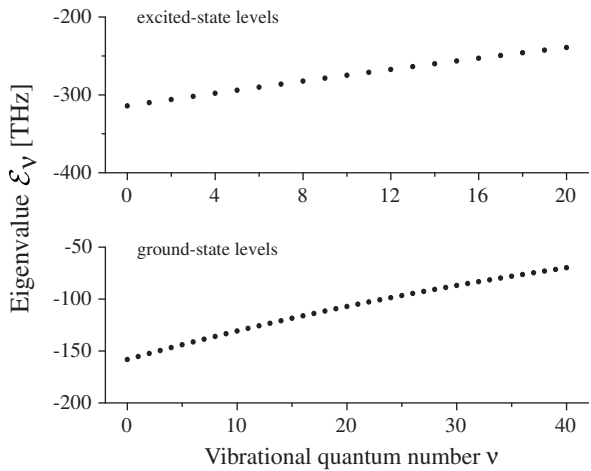}
 \end{center}
\caption{Same as Fig. \ref{fig3} but for the twenty lowest levels
of the excited state and the forty lowest levels of the ground state.
}
\label{fig5}
\end{figure} 

We introduce the combined eigenstates $|a\rangle=|e\rangle\otimes|\varphi_{a}\rangle$
and $|b\rangle=|g\rangle\otimes|\varphi_{b}\rangle$, which are formed from
the internal and translational eigenstates.
The corresponding energies are 
$\hbar\omega_{a}=\hbar\omega_e+\mathcal{E}_{a}$
and $\hbar\omega_{b}=\hbar\omega_g+\mathcal{E}_{b}$. 
Then, we can represent the Hamiltonian (\ref{1}) in the diagonal form
$H_A=\sum_{a}\hbar\omega_{a}|a\rangle\langle a|+
\sum_{b}\hbar\omega_{b}|b\rangle\langle b|$.
We emphasize that the summations over $a$ and $b$ include both the discrete 
($\mathcal{E}_{a,b}<0$) and continuous ($\mathcal{E}_{a,b}>0$) spectra. 
The levels $a$ with $\mathcal{E}_{a}<0$ and the levels $b$ with $\mathcal{E}_{b}<0$
are called the bound (or vibrational) levels of the excited and ground states, respectively. 
In such a state, the atom is bound to the surface. 
It is vibrating, or more exactly, moving back and forth
between the walls formed by the van der Waals potential and the repulsion potential.
The levels $a$ with $\mathcal{E}_{a}>0$ and the levels $b$ with $\mathcal{E}_{b}>0$
are called the free (or continuum) levels of the excited and ground states, respectively.
The center-of-mass wave functions of the bound levels are normalized to unit.
The center-of-mass wave functions of the free levels are normalized per unit energy.
For bound states, the center-of-mass wave functions $\varphi_{a}$ and $\varphi_{b}$ 
can be labeled by the quantum numbers $\nu_a$ and $\nu_b$, respectively. 
For free levels, the conventional sums over $a$ and $b$ must be replaced by the integrals
over $\mathcal{E}_{a}$ and $\mathcal{E}_{b}$, respectively.
 
The center-of-mass wave functions  with energies 
in the range from $-1$ GHz to $-5$ MHz are plotted in Fig. \ref{fig2}.
The center-of-mass eigenvalues in the range from $-1$ GHz to $-20$ kHz are 
displayed in Fig. \ref{fig3}. The mass of  atomic cesium $m=132.9$ a.u. is used
in our calculations. Note that the energy levels in the above range 
are very high compared to the depths  $D_e=316$ THz and $D_g=159.6$ THz 
of the potentials $V_e$ and $V_g$, respectively. Therefore,
the corresponding wave functions are mainly determined by the long-range
van der Waals interaction \cite{Lima}. 
We find that the maximum quantum numbers for the excited- and ground-state potentials 
are 437 and 311, respectively.
When the vibrational quantum number $\nu$ is small, that is, 
when the translational energy level $\mathcal{E}_{\nu}$ is deep, 
the wave function $\varphi_{\nu}$ depends
strongly on the short-range repulsion (see Fig. \ref{fig4}).
For such bound states, 
the center-of-mass wave functions are well confined in the proximity of the surface, 
and the surface-induced frequency shifts of the atomic transitions are substantial compared to the optical frequency $\omega_0=\omega_e-\omega_g\cong352$ THz of the cesium $D_2$ line 
(see Figs. \ref{fig4} and \ref{fig5}).

\subsection{Quantum radiation field and its interaction  with the atom}

We use the formalism of Ref. \cite{Girlanda}
to describe the quantum radiation field in the space with one interface.
We label the modes of the field by the index $\mu=(\omega\mathbf{K}qj)$, where $\omega$ is the mode frequency, $\mathbf{K}=(0,k_y,k_z)$ is the wave-vector component in the surface plane,
$q=s,p$ is the mode polarization, and $j=1,2$ stands for the medium of the input. 
For each mode $\mu=(\omega\mathbf{K}qj)$, the condition $K\leq kn_j$ must be satisfied.
Here, $k=\omega/c$ is the wave number in free space, $n_1>1$ is the refractive index of the dielectric,
and $n_2=1$ is the refractive index of the vacuum.
The mode functions are given, for $x<0$, by \cite{Girlanda}
\begin{eqnarray}
\mathbf{U}_{\omega\mathbf{K}s1}(x)&=&\left(e^{i\beta_{1}x}+
e^{-i\beta_{1}x}r_{12}^s\right)\hat{\mathbf{s}},
\nonumber\\
\mathbf{U}_{\omega\mathbf{K}p1}(x)&=&e^{i\beta_{1}x}\hat{\mathbf{p}}_{1+}+
e^{-i\beta_{1}x}r_{12}^p\hat{\mathbf{p}}_{1-},
\nonumber\\
\mathbf{U}_{\omega\mathbf{K}s2}(x)&=&e^{-i\beta_{1}x} t_{21}^s\hat{\mathbf{s}},
\nonumber\\
\mathbf{U}_{\omega\mathbf{K}p2}(x)&=&e^{-i\beta_{1}x} t_{21}^p\hat{\mathbf{p}}_{1-},
\label{40}
\end{eqnarray} 
and,  for $x>0$, by
\begin{eqnarray}
\mathbf{U}_{\omega\mathbf{K}s1}(x)&=&e^{i\beta_{2}x} t_{12}^s\hat{\mathbf{s}},
\nonumber\\
\mathbf{U}_{\omega\mathbf{K}p1}(x)&=&e^{i\beta_{2}x} t_{12}^p\hat{\mathbf{p}}_{2+},
\nonumber\\
\mathbf{U}_{\omega\mathbf{K}s2}(x)&=&\left(e^{-i\beta_{2}x}+
e^{i\beta_{2}x}r_{21}^s\right)\hat{\mathbf{s}},
\nonumber\\
\mathbf{U}_{\omega\mathbf{K}p2}(x)&=&e^{-i\beta_{2}x}\hat{\mathbf{p}}_{2-}+
e^{i\beta_{2}x}r_{21}^p\hat{\mathbf{p}}_{2+}.
\label{41}
\end{eqnarray}
In Eqs. (\ref{40}) and (\ref{41}), 
the quantity $\beta_j=(k^2n_j^2-K^2)^{1/2}$, 
with $\mathrm{Re}\, \beta_j\geq0$ and $\mathrm{Im}\, \beta_j\geq0$,
is the magnitude of the $x$ component of the wave vector in the medium $j$.
The quantities
$r_{ij}^s=(\beta_{i}-\beta_{j})/(\beta_{i}+\beta_{j})$ and
$t_{ij}^s=2\beta_{i}/(\beta_{i}+\beta_{j})$
are the reflection and transmission Fresnel coefficients for a TE mode,
while the quantities
$r_{ij}^p=(\beta_{i}n_{j}^2-\beta_{j}n_{i}^2)/(\beta_{i}n_{j}^2+\beta_{j}n_{i}^2)$
and $t_{ij}^p=2n_{i}n_{j}\beta_{i}/(\beta_{i}n_{j}^2+\beta_{j}n_{i}^2)$
are the reflection and transmission Fresnel coefficients for a TM mode.
The vector $\hat{\mathbf{s}}=[\hat{\mathbf{K}}\times\hat{\mathbf{x}}]$ 
is the unit vector for the electric field in a TE mode, while the vectors
$\hat{\mathbf{p}}_{j'\pm}=(K\hat{\mathbf{x}}\mp \beta_{j'}\hat{\mathbf{K}})/kn_{j'}$
are the unit vectors for the rightward- and leftward-propagating components of
the electric field in a TM mode in the medium $j'$.

Note that a light beam propagating from the dielectric to the interface may be totally reflected because $n_1>n_2=1$. This phenomenon occurs for the modes $\mu=(\omega\mathbf{K}qj)$ with $j=1$ and $k<K\leq kn_1$. For such a mode, the magnitude of the $x$ component of the wave vector in medium 2 is $\beta_2=i\sqrt{K^2-k^2}$, an imaginary number. 
This mode does not propagate in the $x$ direction in the vacuum side of the interface but decays exponentially. Such a mode is an evanescent mode. 

The total quantized electric field is given by \cite{Girlanda}
\begin{equation}
\mathbf{E}(\mathbf{r})=i\sum_{\mu} 
\frac{k}{4\pi}
\sqrt{\frac{\hbar}{\pi\epsilon_0\beta_{j}}}
\;e^{i\mathbf{K\cdot R}}
\mathbf{U}_{\mu}(x) a_{\mu}e^{-i\omega t}
+\mathrm{H.c.},
\label{45}
\end{equation}
where $a_{\mu}$ is the photon operator for the mode $\mu$, 
and $\sum_{\mu}=\sum_{qj}\int_0^{\infty}d\omega
\int_0^{kn_j}KdK\int_0^{2\pi}d\phi$
is the generalized summation over the modes. Here, $\phi$ is the azimuthal angle of the
vector $\mathbf{K}$ in the $yz$ plane.
The commutation rule for the photon operators is 
$[a_{\mu},a_{\mu'}^\dagger]=
\delta(\omega-\omega')\delta(\mathbf{K}-\mathbf{K}')\delta_{qq'}\delta_{jj'}$.
When dispersion in the region around the frequencies of interest is negligible,
the mode functions $\mathbf{U}_{\mu}$ satisfy the relation
$\int_{-\infty}^\infty dx\, n^2(x) 
\mathbf{U}_{\omega\mathbf{K}qj}^*(x)\mathbf{U}_{\omega'\mathbf{K}q'j'}(x)
=2\pi c^2(\beta_{j}/\omega)\delta(\omega-\omega')\delta_{qq'}\delta_{jj'}$.
Here, $n(x)=n_1$ for $x<0$, and $n(x)=n_2$ for $x>0$.
Hence, we can show that the energy of the field is
$\epsilon_0\int d\mathbf{r}\, n^2(x) 
|\mathbf{E}(\mathbf{r})|^2
=\sum_{\mu}(
a_{\mu}^{\dagger}a_{\mu}+
a_{\mu}a_{\mu}^{\dagger})/2$.
Here, $\int d\mathbf{r}=\int_{-\infty}^{\infty} d x\int_{-\infty}^{\infty} d y\int_{-\infty}^{\infty} d z$ is the integral over the whole space.

We now present the Hamiltonian for the atom--field interaction. 
In the dipole and rotating-wave approximations and in the interaction picture, 
the atom--field interaction  Hamiltonian is 
\begin{equation}
H_{\mathrm{int}}=-i\hbar \sum_{\mu ab}G_{\mu ab}
\sigma_{ab} a_{\mu}
e^{-i(\omega-\omega_{ab})t}+\mbox{H.c.}.
\label{59}
\end{equation}
Here, $\sigma_{ab}=|a\rangle\langle b|$ describes the atomic transition from the level $b$ to the level $a$, 
$\omega_{ab}=\omega_a-\omega_b$ is the angular frequency of the transition, and
\begin{equation}
G_{\mu ab}=\frac{k}{4\pi\sqrt{\pi\epsilon_0\hbar\beta_{j}}}
e^{i\mathbf{K\cdot R}}\,
\langle \varphi_{a}|\mathbf{U}_{\mu} \cdot\mathbf{d}_{eg}| \varphi_{b}\rangle
\label{60}
\end{equation}
is the coupling coefficient. 
In expression (\ref{60}),  $\mathbf{d}_{eg}$ is the dipole moment of the atom,
and $\mathbf{R}=(y,z)$ is the projection of the position vector $\mathbf{r}=(x,y,z)$ of the atom 
onto the interface plane.

\section{Basic equations for spontaneous radiative decay of the atom}
\label{sec:analytical}

We use the mode expansion approach and the Weisskopf--Wigner formalism \cite{Eberly} to derive
the microscopic dynamical equations for spontaneous radiative decay  of 
the atom in the surfaced-induced potential.
For convenience, we introduce the notation 
$\sigma_{\nu\nu'}=|\nu\rangle\langle \nu'|$ and $\omega_{\nu\nu'}=\omega_{\nu}-\omega_{\nu'}$
for the transition operator and the transition frequency, respectively, 
of an arbitrary pair of levels $\nu$ and $\nu'$.
We first study the time evolution of an arbitrary atomic operator $\mathcal{O}$. 
The Heisenberg equation for this operator is
\begin{eqnarray}
\dot{\mathcal{O}}&=&\sum_{\mu ab}(G_{\mu ab}
[\sigma_{ab},\mathcal{O}] a_{\mu}
e^{-i(\omega-\omega_{ab})t}\nonumber\\
&&\mbox{}+G_{\mu ab}^{*}a_{\mu}^{\dagger}[\mathcal{O},\sigma_{ba}]
e^{i(\omega-\omega_{ab})t}.
\label{12n}
\end{eqnarray}
Meanwhile, the Heisenberg equation for the photon operator $a_{\mu}$ is
$\dot{a}_{\mu}=\sum_{ab}G_{\mu ab}^*\sigma_{ba}e^{i(\omega-\omega_{ab})t}$.
Integrating  this equation, we find
\begin{equation}
a_{\mu}(t)=a_{\mu}(t_0)+\sum_{ab}G_{\mu ab}^*\int_{t_0}^t dt'\,
\sigma_{ba}(t')e^{i(\omega-\omega_{ab})t'}.
\label{62}
\end{equation}

We consider the situation where the field is initially in the vacuum state. 
We assume that the evolution time $t-t_0$ and the characteristic atomic lifetime $\tau$ are 
large as compared to the characteristic optical period $T$. 
Under these conditions, since the continuum of the field modes is broadband,
the Markov approximation $\sigma_{ba}(t')=\sigma_{ba}(t)$ can be applied to describe the back
action of the second term in Eq. (\ref{62}) on the atom \cite{Eberly}. 
Under the condition $t-t_0\gg T$, 
we calculate the integral with respect to $t'$ in the limit $t-t_0\to\infty$.  
We set aside the imaginary part of the integral, which describes the frequency shift. 
Such a shift is usually small. 
We can effectively account it  by incorporating into the atomic frequency and the surface--atom potential. 
With the above approximations, we obtain
\begin{equation}
a_{\mu}(t)=a_{\mu}(t_0)+\pi\sum_{ab}G_{\mu ab}^* 
\sigma_{ba}(t)\delta(\omega-\omega_{ab}).
\label{63}
\end{equation}
Inserting Eq. (\ref{63}) into Eq. (\ref{12n}) yields the Heisenberg--Langevin equation
\begin{eqnarray}
\dot{\mathcal{O}}&=&\frac{1}{2}\sum_{aa'bb'}
\big(\gamma_{aa'bb'}[\sigma_{a'b'},\mathcal{O}] \sigma_{ba}e^{-i(\omega_{ab}-\omega_{a'b'})t}
\nonumber\\&&\mbox{}
+\gamma_{aa'bb'}^*\sigma_{ab}[\mathcal{O},\sigma_{b'a'}] e^{i(\omega_{ab}-\omega_{a'b'})t}\big)
+\xi_{\mathcal{O}}.
\label{13n}
\end{eqnarray}
Here, 
\begin{equation}
\gamma_{aa'bb'}=2\pi\sum_{\mu}G_{\mu ab}^*G_{\mu a'b'}\delta(\omega-\omega_{ab}) 
\label{66n}
\end{equation}
are the decay coefficients and $\xi_{\mathcal{O}}$ is the noise operator. 
We emphasize that Eq. (\ref{13n}) can be applied to any atomic operators.

We now examine the time evolution of the reduced density 
operator $\rho$ of the atomic system.
We multiply Eq. (\ref{13n}) with $\rho(0)$, take the trace of the result,
use the relation $\mathrm{Tr}[\mathcal{O}(t)\rho(0)]=\mathrm{Tr}[\mathcal{O}(0)\rho(t)]$, 
transform to arrange the operator $\mathcal{O}(0)$ at the first position in each operator product, and eliminate $\mathcal{O}(0)$. Then, we obtain the Liouville equation
\begin{eqnarray}
\dot{\rho}&=&\frac{1}{2}\sum_{aa'bb'}
\big[\gamma_{aa'bb'}(\sigma_{ba}\rho\sigma_{a'b'}-\delta_{bb'} \sigma_{a'a}\rho)
e^{-i(\omega_{ab}-\omega_{a'b'})t}
\nonumber\\&&\mbox{}
+\gamma_{aa'bb'}^*(\sigma_{b'a'}\rho\sigma_{ab}-\delta_{bb'}\rho\sigma_{aa'}) 
e^{i(\omega_{ab}-\omega_{a'b'})t}\big].
\label{13a}
\end{eqnarray}
For the matrix elements $\rho_{\nu\nu'}=\langle\nu|\rho|\nu'\rangle$, the above equations yields
\begin{eqnarray}
\dot{\rho}_{aa'}&=&
-\frac{1}{2}\sum_{a''}(\gamma_{a''a}e^{i\omega_{aa''}t}\rho_{a''a'}
+\gamma_{a''a'}^*e^{i\omega_{a''a'}t}\rho_{aa''}),
\nonumber\\
\dot{\rho}_{ab}&=&
-\frac{1}{2}\sum_{a'}\gamma_{a'a}e^{i\omega_{aa'}t}\rho_{a'b},
\nonumber\\
\dot{\rho}_{bb'}&=&\frac{1}{2}\sum_{aa'}(\gamma_{aa'bb'}+\gamma_{a'ab'b}^*)
e^{i(\omega_{bb'}-\omega_{aa'})t}{\rho}_{aa'},
\label{66a}
\end{eqnarray}
where $\gamma_{aa'}=\sum_b\gamma_{aa'bb}
=2\pi\sum_{\mu b}G_{\mu ab}^*G_{\mu a'b}\delta(\omega-\omega_{ab})$.
Equations (\ref{66a}) show that
the spontaneous decay rate for a transition $a\to b$ is given by
$\gamma_{ab}=\gamma_{aabb}=2\pi\sum_{\mu}|G_{\mu ab}|^2\delta(\omega-\omega_{ab})$.
Meanwhile, the total decay rate for an excited level $a$ is
$\gamma_{a}=\gamma_{aa}=2\pi\sum_{\mu b}|G_{\mu ab}|^2\delta(\omega-\omega_{ab})$.
It is obvious that $\gamma_{a}=\sum_{b}\gamma_{ab}$. 

Equations (\ref{66a}) correspond to the incoherent radiative decay part of the master equation for an atom interacting with external fields in the proximity of a surface. 
A set of more general equations is presented in Appendix \ref{sec:driven}
for the case where the atom is driven by classical coherent plane-wave fields.
We note that the terms in Eqs. (\ref{66a}) are different from and much more complicated than the conventional phenomenological decay terms \cite{Boyd}.
The unusual structure of Eqs. (\ref{66a}) results from 
the complexity of the translational levels of the atom in the surface-induced potential.
Similar structures are also observed in the case of 
a multilevel alkali atom \cite{cesium decay,ChangMinogin}.

\section{spontaneous radiative decay rates}
\label{sec:rates}

In the half-space $x>0$, where the atom is restricted to, 
the mode functions of the radiation field are described by expressions (\ref{41}).
We insert these expressions into Eq. (\ref{60}) and then the result into Eq. (\ref{66n}). 
We perform the summation with respect to the mode index $\mu$.
Then, in the case where the dipole moment $\mathbf{d}_{eg}$ is perpendicular to the interface, we obtain
\begin{eqnarray}
\gamma_{aa'bb'}^{\perp}&=&\frac{3}{2}\gamma_f(\omega_{ab})
\Bigg\{\int\limits_0^1 \mathrm{Re}\Big[(1-\xi^2)F_{ab}(\xi k_{ab})F_{a'b'}^*(\xi k_{ab})
\nonumber\\&&\mbox{}
+r_{\perp}(\xi)F_{ab}(\xi k_{ab})F_{a'b'}(\xi k_{ab})\Big]d\xi
\nonumber\\&&\mbox{}
+\int\limits_0^{\sqrt{n_1^2-1}}
T_{\perp}(\xi)
I_{ab}(\xi k_{ab})I_{a'b'}(\xi k_{ab})\,d\xi\Bigg\},
\label{74}
\end{eqnarray}
and, in the case where the dipole moment $\mathbf{d}_{eg}$ is parallel to the interface, we find
\begin{eqnarray}
\gamma_{aa'bb'}^{\parallel}&=&\frac{3}{4}\gamma_f(\omega_{ab})
\Bigg\{\int\limits_0^1 \mathrm{Re}\Big[(1+\xi^2)F_{ab}(\xi k_{ab})F_{a'b'}^*(\xi k_{ab})
\nonumber\\&&\mbox{}
+r_{\parallel}(\xi)F_{ab}(\xi k_{ab})F_{a'b'}(\xi k_{ab})\Big]d\xi
\nonumber\\&&\mbox{}
+\int\limits_0^{\sqrt{n_1^2-1}}
T_{\parallel}(\xi)
I_{ab}(\xi k_{ab})I_{a'b'}(\xi k_{ab})\,d\xi\Bigg\}.
\label{75}
\end{eqnarray}
Here we have introduced the notation 
\begin{eqnarray}
r_{\perp}(\xi)&=&(1-\xi^2)\frac{n_{1}^2\xi-\sqrt{n_1^2-1+\xi^2}}
{n_{1}^2\xi+\sqrt{n_1^2-1+\xi^2}},
\nonumber\\
r_{\parallel}(\xi)&=&
\frac{\xi-\sqrt{n_1^2-1+\xi^2}}{\xi+\sqrt{n_1^2-1+\xi^2}}
-\xi^2\frac{n_1^2\xi-\sqrt{n_1^2-1+\xi^2}}{n_1^2\xi+\sqrt{n_1^2-1+\xi^2}},
\nonumber\\ 
\label{76}
\end{eqnarray}
and
\begin{eqnarray}
T_{\perp}(\xi)&=&\frac{2n_1^2}{n_1^2-1}\,
\frac{\sqrt{n_1^2-1-\xi^2}}{(n_1^2+1)\xi^2+1}\,\xi(1+\xi^2),
\nonumber\\ 
T_{\parallel}(\xi)&=&
\frac{2}{n_1^2-1}
\left(1+\frac{n_1^2\xi^2}{(n_1^2+1)\xi^2+1}\right)
\xi\sqrt{n_1^2-1-\xi^2}.
\nonumber\\
\label{77}
\end{eqnarray}
The quantity  
\begin{equation}
\gamma_f(\omega)=\frac{d_{eg}^2\omega^3}{3\pi\epsilon_0\hbar c^3}
\label{54}
\end{equation}
is the natural linewidth of a two-level atom with the transition frequency 
$\omega$ and the dipole moment $d_{eg}$.
The matrix elements
\begin{eqnarray}
F_{ab}(\beta)&=&\langle \varphi_{a}|e^{i\beta x}| \varphi_{b}\rangle,
\nonumber\\ 
I_{ab}(\beta)&=&\langle \varphi_{a}|e^{-\beta x}| \varphi_{b}\rangle
\label{6}
\end{eqnarray}
for the transitions between the translational eigenstates $|\varphi_{a}\rangle$
and $|\varphi_{b}\rangle$ have been introduced. 
In deriving Eqs. (\ref{74}) and (\ref{75}) we have changed
the integration variable $K$ to $\xi=\sqrt{|1-\kappa^2|}$. Here the parameter $\kappa=K/k_{ab}$ with 
$k_{ab}=\omega_{ab}/c$ has been introduced. 

The functions $r_{\perp}$ and $r_{\parallel}$
are related to the reflection coefficients $r_{21}^s=(\xi-\eta)/(\xi+\eta)$ 
and $r_{21}^p=(n_1^2\xi-\eta)/(n_1^2\xi+\eta)$ as given by
$r_{\perp}=\kappa^2 r_{21}^p$ and 
$r_{\parallel}=r_{21}^s-\xi^2 r_{21}^p$.
Here, $\kappa=\sqrt{1-\xi^2}$ and $\eta=\sqrt{n_1^2-1+\xi^2}$.
Hence, the terms that contain
$r_{\perp}$ and $r_{\parallel}$ in Eqs. (\ref{74}) and (\ref{75}) 
are the results of the interference between the emitted and reflected fields.
Meanwhile, the functions $T_{\perp}$ and $T_{\parallel}$
are related to the transmission coefficients $t_{12}^s=2\eta/(\eta+i\xi)$ 
and $t_{12}^p=2n_1\eta/(\eta+in_1^2\xi)$ of the evanescent modes as given by 
$T_{\perp}=(\xi/2\eta) \kappa^2|t_{12}^p|^2$ and 
$T_{\parallel}=(\xi/2\eta)(|t_{12}^s|^2+\xi^2 |t_{12}^p|^2)$.
Here, $\kappa=\sqrt{1+\xi^2}$ and $\eta=\sqrt{n_1^2-1-\xi^2}$.
Hence, the terms that contain
$T_{\perp}$ and $T_{\parallel}$ in Eqs. (\ref{74}) and (\ref{75}) 
are the results of the emission into the evanescent modes. 

The matrix elements $F_{ab}(\beta)$ and $I_{ab}(\beta)$, defined by Eqs. (\ref{6}), depend on the overlap between the translational wave functions $\varphi_a$ and $\varphi_b$. 
The factors  $|F_{ab}(0)|^2$ and $|I_{ab}(0)|^2$, 
with the argument $\beta=0$, are the same and equal to the
Franck--Condon factors $|\langle \varphi_{a}|\varphi_{b}\rangle|^2$.
The exponential factors $e^{i\beta x}$ and $e^{-\beta x}$ in Eqs. (\ref{6})
take into account the momentum of the photon emitted or absorbed by the atom 
(see Appendix \ref{sec:driven}). 
The factor $e^{i\beta x}$ corresponds to the case where the mode is a propagating mode, and 
the factor $e^{-\beta x}$ corresponds to the case where the mode is an evanescent mode. 
Thus, $F_{ab}(\beta)$  characterizes the strength of the translational transition  
with the participation of a propagating-mode photon,
and $I_{ab}(\beta)$ corresponds to the case of an evanescent-mode photon. 

We assume that the orientation of the atomic dipole moment is completely random.  
In this case, the averaged decay parameters are given by 
$\gamma_{aa'bb'}=(\gamma_{aa'bb'}^{\perp}+2\gamma_{aa'bb'}^{\parallel})/3$. 
Using Eqs. (\ref{74}) and (\ref{75}), we find 
\begin{eqnarray}
\lefteqn{
\gamma_{aa'bb'}=\gamma_f(\omega_{ab})
\Bigg\{\int\limits_0^1 \mathrm{Re}\Big\{ F_{ab}(\xi k_{ab})F_{a'b'}^*(\xi k_{ab})}
\nonumber\\&&\mbox{}
+\frac{r_{\perp}(\xi)+r_{\parallel}(\xi)}{2} 
F_{ab}(\xi k_{ab})F_{a'b'}(\xi k_{ab})\Big\}d\xi
\nonumber\\&&\mbox{}
+\int\limits_0^{\sqrt{n_1^2-1}}
\frac{T_{\perp}(\xi)+T_{\parallel}(\xi)}{2}
I_{ab}(\xi k_{ab})I_{a'b'}(\xi k_{ab})\,d\xi\Bigg\}.\qquad
\label{81}
\end{eqnarray}
We note that all the decay coefficients are real quantities.

When we set $a'=a$ and $b'=b$ in Eq. (\ref{81}), 
we obtain the following expression for the spontaneous transition rate: 
\begin{eqnarray}
\gamma_{ab}&=&\gamma_f(\omega_{ab})
\Bigg\{\int\limits_0^1 \Big\{|F_{ab}(\xi k_{ab})|^2
\nonumber\\&&\mbox{}
+\frac{r_{\perp}(\xi)+r_{\parallel}(\xi)}{2}\mathrm{Re}[F_{ab}^2(\xi k_{ab})]\Big\}d\xi
\nonumber\\&&\mbox{}
+\int\limits_0^{\sqrt{n_1^2-1}}
\frac{T_{\perp}(\xi)+T_{\parallel}(\xi)}{2}I_{ab}^2(\xi k_{ab})\,d\xi\Bigg\}.
\label{82}
\end{eqnarray}
Equation (\ref{82}) shows that the effects of the dielectric 
on the spontaneous transition rate 
$\gamma_{ab}$ appear through (1) the shift of the transition frequency $\omega_{ab}$,
(2) the overlap $|F_{ab}|^2$ between the center-of-mass wave functions,
(3) the interference between the emitted and reflected fields, 
and (4) the transmission into the evanescent modes.

The cross decay coefficient $\gamma_{aa'}=\sum_{b}\gamma_{aa'bb}$
is the sum of $\gamma_{aa'bb}$  over the ground-state levels $b$.  
The summation can be simplified if,
for each fixed index $a$, the overlapping factors $F_{ab}$ and $I_{ab}$
are substantial only in a small region
of $b$. In this case,  the transition frequency $\omega_{ab}$ in expression (\ref{81})
can be replaced by an effective frequency $\omega_{a\bar{b}}$ that does not depend on $b$. 
We use this approximation, set $b'=b$, and sum up the result with respect to $b$.
Then, with the help of the
relation $\sum_b |\varphi_b\rangle\langle\varphi_b|=1$, we obtain
\begin{equation}
\gamma_{aa'}=\langle \varphi_{a}|\gamma_x(\omega_{a\bar{b}})| \varphi_{a'}\rangle.
\label{70}
\end{equation}
Here,
\begin{eqnarray}
\gamma_x(\omega)&=&\gamma_f(\omega)
\Bigg\{1+\int\limits_0^1 
\frac{r_{\perp}(\xi)+r_{\parallel}(\xi)}{2}\cos(2\xi kx)d\xi
\nonumber\\&&\mbox{}
+\int\limits_0^{\sqrt{n_1^2-1}}
\frac{T_{\perp}(\xi)+T_{\parallel}(\xi)}{2}e^{-2\xi kx}d\xi\Bigg\}
\label{69}
\end{eqnarray}
is the spontaneous emission rate of a two-level atom with frequency $\omega$, 
being at rest in the vicinity of the dielectric \cite{Agarwal,Wylie}.
In order to find the linewidth $\gamma_a$, we set $a=a'$ in Eq. (\ref{70}). Then, we obtain
\begin{equation}
\gamma_{a}=\langle \varphi_{a}|\gamma_x(\omega_{a\bar{b}})| \varphi_{a}\rangle.
\label{72}
\end{equation}
The above equation means that the  radiative linewidth $\gamma_a$ can be approximately considered as the average of the effective-frequency rest-atom decay rate $\gamma_x(\omega_{a\bar{b}})$ with respect to the center-of-mass wave function $\varphi_a(x)$.

We note that, in the case where the refractive index $n_1$ of the dielectric is not large, 
the reflection of light from the interface
and the emission of light into the evanescent modes are not strong. 
In this case,
$\gamma_x(\omega)$ with a fixed argument $\omega$ 
is a slowly varying function of $x$. Hence, for $a\not=a'$,
since $|a\rangle$ and $|a'\rangle$ are orthogonal to each other,
the quantity $\gamma_{aa'}$  is usually small compared to $\gamma_a$ and $\gamma_{a'}$.   

We emphasize  that the first integral in Eq. (\ref{69}) results from the interference
between the emitted and reflected fields.
The second integral in Eq. (\ref{69}) results from the emission
into the evanescent modes.
We note that expression (\ref{69}) for $\gamma_x$ is in full agreement with the results of the linear-response formalism \cite{Agarwal,Wylie}. 
However, the results of Refs.  \cite{Agarwal,Wylie} are written in a different form
that is difficult to identify the origins and physical meanings of the contributions. 

If we neglect reflection from the interface
and emission into the evanescent modes and limit ourselves to considering 
only levels with negligible surfaced-induced frequency shifts, 
then Eq. (\ref{81}) reduces to 
\begin{equation}
\gamma_{aa'bb'}=\gamma_0f_{aa'bb'},
\label{83}
\end{equation}
where $\gamma_0=\gamma_f(\omega_0)$, $\omega_0=\omega_e-\omega_g$, and 
\begin{eqnarray}
\lefteqn{
f_{aa'bb'}=\frac{1}{2}
\int_{-1}^{1}d\xi\,
\langle\varphi_a|e^{i\xi k_0x}|\varphi_b\rangle
\langle\varphi_{b'}|e^{-i\xi k_0x}|\varphi_{a'}\rangle}
\nonumber\\
&=&\int\limits_{-\infty}^{\infty}\!\!\int\limits_{-\infty}^{\infty} dxdx'
\frac{\sin k_0(x-x')}{k_0(x-x')}\varphi_a(x)\varphi_b(x)\varphi_{a'}(x')\varphi_{b'}(x').
\nonumber\\
\label{84}
\end{eqnarray} 
Setting $b'=b$ in expressions (\ref{83}) and (\ref{84}) and 
summing up the resultant expressions over $b$, we find 
$\gamma_{aa'}=\gamma_0\langle a|a'\rangle$.

\section{Numerical results}
\label{sec:numerical}

In what follows,  we present the results of our numerical calculations for the spontaneous radiative decay characteristics of the atom. As stated in Sec. \ref{sec:model}, we use the parameters of fused silica, for the dielectric, and the parameters of atomic cesium with the $D_2$ line, for the atom. The refractive index of the medium is $n_1=1.45$. 
The wavelength of the $D_2$ line of atomic cesium in free space is $\lambda_0=852$ nm. 
Other parameters are given in Sec. \ref{sec:model}.

\begin{figure}[tbh]
\begin{center}
  \includegraphics{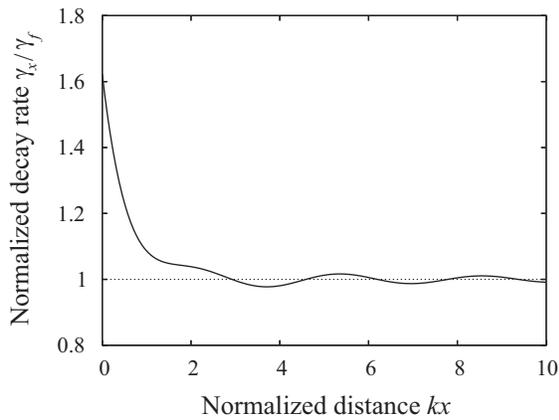}
 \end{center}
\caption{Normalized decay rate $\gamma_x(\omega)/\gamma_f(\omega)$ of a two-level atom rested at a point in the vicinity of a semi-infinite dielectric medium 
as a function of the normalized surface--atom distance $kx$. 
The refractive index of the medium is $n_1=1.45$. 
}
\label{fig6}
\end{figure} 
  
We first consider the case where the center-of-mass motion of the atom can be neglected.
We plot in Fig. \ref{fig6} the normalized decay rate $\gamma_x(\omega)/\gamma_f(\omega)$ as a function of the normalized surface--atom distance $kx$. 
The figure shows that $\gamma_x(\omega)/\gamma_f(\omega)$
varies slowly with $kx$. 
We observe not only enhancement, $\gamma_x/\gamma_f>1$, 
but also inhibition, $\gamma_x/\gamma_f<1$, 
of spontaneous emission, depending on the position of the atom. 
Such changes are quantum electrodynamic effects resulting from the modifications of the field modes in the presence of the dielectric \cite{Agarwal,Wylie}.
The maximum value of $\gamma_x(\omega)/\gamma_f(\omega)$ is about $1.6$, achieved at $kx=0$.
Such a quantum enhancement is moderate. It is not dramatic. The reason is that the refractive index of silica $n_1=1.45$ is not large.
We observe small oscillations in $\gamma_x(\omega)/\gamma_f(\omega)$ as $kx$ increases. 
Such oscillations are due to the interference between the emitted and reflected fields.

\begin{figure}[tbh]
\begin{center}
  \includegraphics{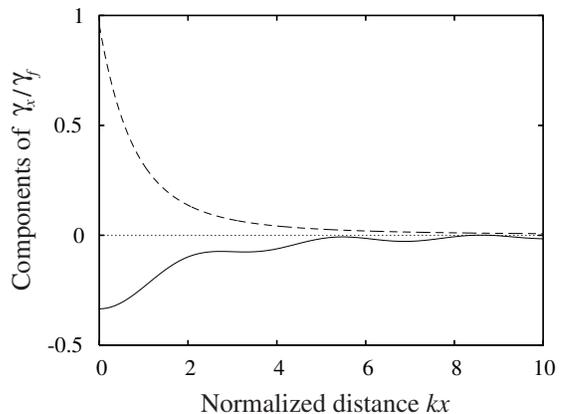}
 \end{center}
\caption{Contributions of the first (solid line) and
second (dashed line) integrals in Eq. (\ref{69}) to
the normalized decay rate $\gamma_x(\omega)/\gamma_f(\omega)$ of Fig. \ref{fig6}.
}
\label{fig7}
\end{figure} 

The deviation of $\gamma_x(\omega)/\gamma_f(\omega)$ from unity is caused by the interference between the emitted and reflected fields and by the emission into the evanescent modes, 
which are expressed by the first and second integrals in Eq. (\ref{69}), respectively. 
We calculate these integrals separately and plot the results in Fig. \ref{fig7}. 
The solid curve of the figure shows that, for $kx\leq 8$, 
the first integral is negative. Thus, in the close vicinity of the interface, the interference between the emitted and reflected fields is destructive and hence tends to reduce the spontaneous decay rate. The solid curve of the figure also shows that, in the region of large $kx$, the first integral can become positive, that is, the interference can be constructive, depending on the position of the atom. 
Meanwhile, the dashed curve of the figure shows that the second integral 
in Eq. (\ref{69}) is always positive. 
Thus, the emission into the evanescent modes always enhances the spontaneous decay rate.
Such an enhancement is due to the existence and confinement of the evanescent modes.

\begin{figure}[tbh]
\begin{center}
  \includegraphics{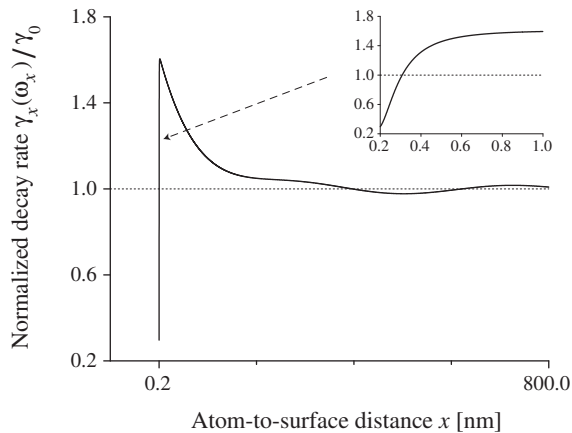}
 \end{center}
\caption{Normalized decay rate $\gamma_x(\omega_x)/\gamma_0$ 
of a two-level atom with a space-dependent frequency, rested at a point in the vicinity of a semi-infinite medium. 
The transition frequency $\omega_x$ is shifted from 
the free-space transition frequency $\omega_0$ by the surface--atom interaction.
The parameters of the potentials are as in Fig. \ref{fig1}.
The refractive index of the medium is $n_1=1.45$.
The wavelength of the transition of the atom in free space is $\lambda_0=852$ nm. 
}
\label{fig8}
\end{figure} 

Due to the surface--atom potentials $V_e$ and $V_g$, the transition frequency of the atom varies in space as given by
$\omega=\omega_x=\omega_0+[V_e(x)-V_g(x)]/\hbar$. Consequently, when we take into account the frequency shift, the decay rate of the atom is $\gamma_x(\omega_x)$. 
The spatial dependence of 
$\gamma_x(\omega_x)$ is determined not only by the change in the mode structure but also by the change in the atomic transition frequency $\omega_x$. We plot the spatial dependence of 
$\gamma_x(\omega_x)$ in Fig. \ref{fig8}. 
The figure and the inset show that, when $x$ decreases from 1 nm to 0.2 nm, the decay rate drops quickly from its peak value. The reason is that, in this region of space, the surface-induced frequency shift of the transition is negative (red shift) and substantial.
Such a shift of $\omega_x$ reduces the natural linewidth $\gamma_f(\omega_x)$, 
see Eq. (\ref{54}), and hence reduces the decay rate $\gamma_x(\omega_x)$, 
see Eq. (\ref{69}).  
The effects of the transition frequency shift and the quantum enhancement  
on the decay are substantial in the regions 
$x\lesssim [(C_{3e}-C_{3g})/\hbar\omega_0]^{1/3}=0.16$ nm and $x\lesssim\lambda_0=852$ nm, respectively.
The two scales are quite different from each other, namely, 
$[(C_{3e}-C_{3g})/\hbar\omega_0]^{1/3}\ll\lambda_0$.
In the region $x>1 $ nm, the frequency shift of the transition is small and hence the spatial dependence of the decay rate is mainly determined by the mode structure of the field.
Meanwhile, in the region from 1 nm to 0.2 nm, the effect of the frequency shift is dominant.  
We note that, in the region $x<0.2$ nm, 
which is not shown in the figure, the transition frequency shift 
may become positive leading to an enhancement of the decay rate.
In this region, the potential slope is steep, the force is large, and consequently 
the consideration of the center-of-mass motion of the atom must be included.
Due to the lack of quantitative information about the silica--cesium repulsive potential and due to the need to include the center-of-mass motion of the atom, we do not plot the decay rate $\gamma_x(\omega_x)$ in the region $x<0.2$ nm. The cutoff value of $0.2$ nm is chosen because it is close to the position $x_m=0.19$ nm 
of the minima of the potentials $V_e$ and $V_g$. 

\begin{figure}[tbh]
\begin{center}
  \includegraphics{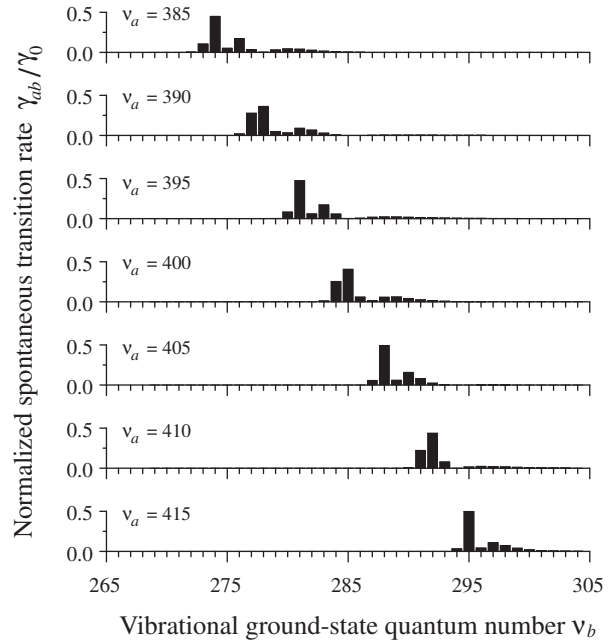}
 \end{center}
\caption{
Spontaneous transition rates $\gamma_{ab}$ for levels with large vibrational quantum numbers. The rates are normalized to the free-space spontaneous decay rate $\gamma_0$.
The parameters used are as in Figs. \ref{fig1}, \ref{fig2}, and \ref{fig8}.
}
\label{fig9}
\end{figure} 

In the close vicinity of the surface, we must take into account the center-of-mass motion of the atom. To do this in a fully quantum treatment, 
we must consider the sets of combined states 
$\{|a\rangle\}$ and $\{|b\rangle\}$ instead of the internal states $|e\rangle$ and $|g\rangle$, respectively. We plot in Fig. \ref{fig9} the spontaneous transition rate $\gamma_{ab}$ between shallow levels $a$ and $b$, with large vibrational quantum numbers. 
The figure shows that each upper level $a$ can be substantially coupled to one or several lower levels $b$. In general, $\gamma_{ab}$ varies unevenly with increasing 
$\nu_a$ or $\nu_b$. The  values of $\nu_b$ for which $\gamma_{ab}$ is substantial tend to become larger when $\nu_a$ increases.
Such features are due to the fact that $\gamma_{ab}$ substantially depends on the overlap
and interference between the center-of-mass wave functions $\varphi_a$ and $\varphi_b$. 
It is clear from Figs. \ref{fig3} and \ref{fig9} that the frequency shifts
of the significant transitions between the bound levels $a$ and $b$ with large vibrational quantum numbers are small compared to the free-space optical frequency $\omega_0$.

\begin{figure}[tbh]
\begin{center}
  \includegraphics{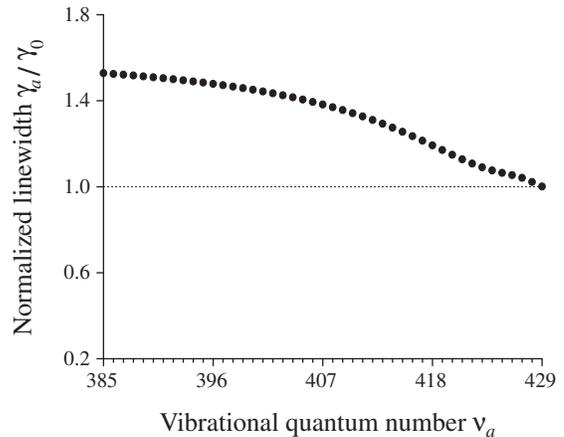}
 \end{center}
\caption{Radiative linewidths $\gamma_a$ of bound excited-state levels 
with large vibrational quantum numbers. 
The linewidths are normalized to the free-space spontaneous decay rate $\gamma_0$.
The parameters used are as in Figs. \ref{fig1}, \ref{fig2}, and \ref{fig8}.
}
\label{fig10}
\end{figure} 

We use the approximate equation (\ref{72}) to calculate
the linewidths $\gamma_a$ of shallow bound excited-state levels $a$, with 
large quantum numbers $\nu_a$, in the range from 385 to 429, and 
plot the results in Fig. \ref{fig10}. 
The figure  shows that $\gamma_a/\gamma_0$ increases slowly from
1 to  1.53 as $\nu_a$ reduces from 429 to 385. 
The observed enhancement of $\gamma_a$
is the average of the quantum enhancement of 
the rest-atom decay rate $\gamma_x(\omega_{a\bar{b}})$ with respect to the center-of-mass wave function $\varphi_a(x)$ [see Eq. (\ref{72})]. When $\nu_a$ is large enough,
the effect of the surface--atom interaction on the transition frequency is small,
that is, $\omega_{a\bar{b}}\cong\omega_0$. In this case, the magnitude of $\gamma_a$ is mainly determined by the mode structure.
When $\nu_a$ is not too large, the center-of-mass wave function $\varphi_a(x)$ is confined in a small spatial region close to the surface and, hence, the quantum enhancement of $\gamma_a$ is observed.
The spatial spread of $\varphi_a(x)$ increases with increasing $\nu_a$.
This explains why $\gamma_a$ reduces with increasing $\nu_a$ in Fig. \ref{fig10}. 
When $\nu_a$ is very large, the quantum enhancement is negligible and, 
hence, we have $\gamma_a\cong\gamma_0$.

\begin{figure}[tbh]
\begin{center}
  \includegraphics{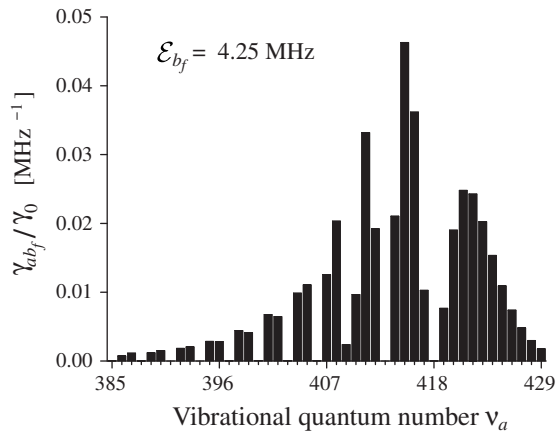}
 \end{center}
\caption{Normalized density $\gamma_{ab_f}/\gamma_0$ for the rate of spontaneous radiative decay from a bound excited-state level $a$ to a free ground-state level $b_f$ as a function of the vibrational quantum number $\nu_a$.
The parameters used are as in Figs. \ref{fig1}, \ref{fig2}, and \ref{fig8}.
}
\label{fig11}
\end{figure} 

The spectra of translational levels of ground and excited states include not only bound
levels but also free levels. Both types of levels  
can be coupled to each other. Since the center-of-mass wave functions 
of free excited-state levels $a_f$ and free ground-state levels $b_f$ are normalized
per unit energy, the quantities $\gamma_{ab_f}$, $\gamma_{a_fb}$, and $\gamma_{a_fb_f}$ 
are the densities of the spontaneous decay rates for the transitions $a\to b_f$,
$a_f\to b$, and $a_f\to b_f$, respectively. The knowledge of these quantities is required
for  the study of the dynamical and spectroscopic characteristics of moving atoms. For example, in order to calculate the excitation spectrum of an atom initially prepared in a free ground-state level $b_f$, we need to know the linewidths of the transitions $a\to b_f$.
In this connection, we plot in Fig. \ref{fig11} the normalized density
$\gamma_{ab_f}/\gamma_0$ for the rate of spontaneous decay from a bound excited-state level $a$ to a free ground-state level $b_f$ as a function of 
the vibrational quantum number $\nu_a$.

\begin{figure}[tbh]
\begin{center}
  \includegraphics{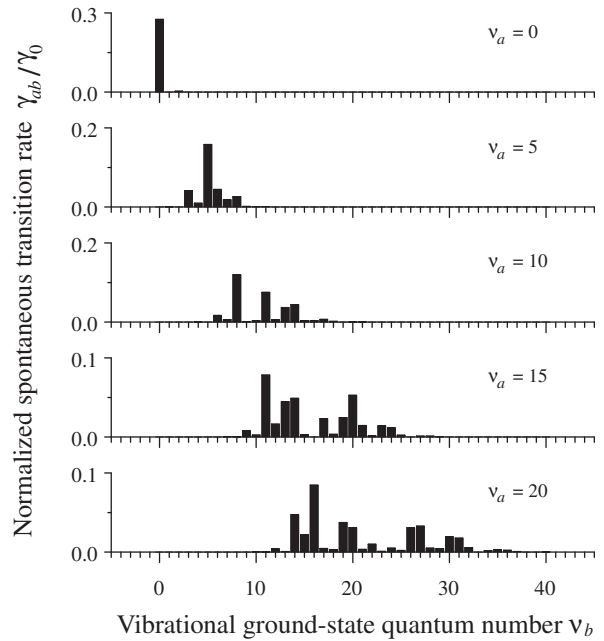}
 \end{center}
\caption{Same as Fig. \ref{fig8} but for levels 
with small vibrational quantum numbers.
}
\label{fig12}
\end{figure}

\begin{figure}[tbh]
\begin{center}
  \includegraphics{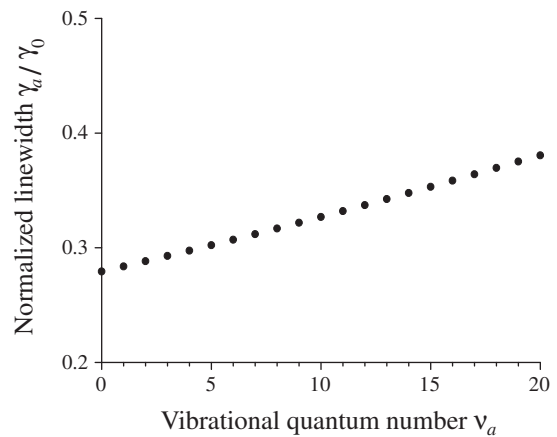}
 \end{center}
\caption{Same as Fig. \ref{fig10} but for levels 
with small vibrational quantum numbers.
}
\label{fig13}
\end{figure} 

The effect of the surface-induced shifts of the transition frequencies on the transition rates and linewidths becomes substantial when the translational eigenfunctions of the atom are close to the surface, that is, when the levels are deep.  
We plot in Figs. \ref{fig12} and \ref{fig13} the spontaneous transition rates $\gamma_{ab}$ and the linewidths $\gamma_a$, respectively, 
for deep bound levels, with small vibrational quantum numbers.
Our additional numerical calculations (not presented here) show that the transitions from bound excited-state levels with small $\nu_a$ to free ground-state levels are negligible. 
Due to this fact, the sum of the values of the individual 
transition rates $\gamma_{ab}$ presented
in each fixed row of Fig. \ref{fig12} almost coincides with the corresponding value of the linewidth $\gamma_a$ in Fig. \ref{fig13}. According to Fig. \ref{fig13}, 
in the region of small $\nu_a$, we have $\gamma_a<\gamma_0$, and $\gamma_a$ reduces with decreasing $\nu_a$. In this region, 
the surface-induced transition frequency shift is negative and substantial, leading to a decrease in $\omega_{ab}$ and, consequently, in $\gamma_f(\omega_{ab})$, $\gamma_{ab}$, and $\gamma_a$. The minimum value of the radiative linewidth $\gamma_a$ is observed for $\nu_a=0$ and is about $0.28\gamma_0$. This value is substantially (about 3.6 times) smaller than the natural linewidth $\gamma_0$ of the atom. 
We note that, when the atomic center-of-mass states are close to the surface and
the surface temperature is high, the decay of the atom due to the interaction with 
the phonons of the solid may become important \cite{Oria2006}.
However, the study of the decay due to phonons is beyond the scope of this paper.

\section{Conclusions}
\label{sec:summary}

We have studied spontaneous radiative decay of translational levels of an atom in the vicinity of a semi-infinite dielectric. We have systematically derived the microscopic dynamical equations for the spontaneous decay process. We have calculated analytically and numerically the spontaneous transition rates between the bound excited and bound ground states and 
the radiative linewidths of the bound excited states. 
We have shown that the effects of the dielectric 
on the spontaneous transition rates and radiative linewidths  
appear through (1) the shift of the transition frequency,
(2) the overlap between the center-of-mass wave functions,
(3) the interference between the emitted and reflected fields, 
and (4) the transmission into the evanescent modes.
Our numerical calculations for the silica--cesium interaction  
have demonstrated that the radiative linewidths of 
the bound excited levels with large enough but not too large vibrational quantum numbers 
are moderately enhanced by the emission into the evanescent modes and those for the deep bound levels are substantially reduced by the surface-induced red shift of the transition frequency.
We believe that the radiative longevity of deep bound states 
can find important applications for atom trapping, quantum information, and 
atom and photonic nano-devices. 
We emphasize that we have studied in this paper only spontaneous radiative decay.
The problem of decay due to phonons of the solid is beyond the scope
of our paper.

\begin{acknowledgments}
We thank M. Ori\'{a}, S. Dutta Gupta, and P. N. Melentiev for fruitful discussions.
This work was carried out under the 21st Century COE program on ``Coherent Optical Science''.
\end{acknowledgments}

\appendix

\section{Parameters of the silica--cesium potential}
\label{sec:potential}

Consider the potential 
$V(x)=A e^{-\alpha x}-C_{3}/x^3$.
Under the condition 
$\alpha^3C_3/A<256/3e^{4}$,
this potential has  a peak at $x_p=\xi_p/\alpha$ and
a local minimum at $x_m=\xi_m/\alpha$, where $\xi_p<4$ and $\xi_m>4$  
are the solutions to the equation $\xi^4e^{-\xi}=3\alpha^3C_3/A$. 
The peak value $V(x_p)$ should be positive and large enough to create a steep slope for the potential $V(x)$ in the small interval $(x_p,x_m)$. Such a slope leads to a substantial repulsive force on the atom in the close vicinity of the surface. We discard the region $x<x_p$, assuming that the atom is outside and cannot penetrate into this region of space. 

The depth of the potential is defined by
$D=-V(x_m)=-A e^{-\alpha x_m}+C_3/x_m^3$. We find the relation 
\begin{equation}
A=\frac{3D}{\alpha x_m-3}e^{\alpha x_m}. 
\label{26}
\end{equation}
In terms of the parameter $x_a=\sqrt[3]{C_3/D}$, we have
\begin{equation}
\alpha=\frac{3}{x_m[1-(x_m/x_a)^3]}.
\label{28}
\end{equation}
Equations (\ref{26}) and (\ref{28}) allow us to determine the parameters $A$ and $\alpha$
from the parameters $C_3$, $D$, and $x_m$. Since $\alpha>0$  and $\alpha x_m>4$, the condition
\begin{equation}
\frac{x_a}{\sqrt[3]{4}}<x_m<x_a
\label{29}
\end{equation}
must be satisfied.

We  use a few available theoretical and experimental data to  estimate the silica--cesium potential parameters. According to Ref. \cite{Safronova}, the van der Waals coefficient for the interaction between a ground-state cesium atom and a perfect metal surface is 
$C_3^{(\mathrm{metal})}=4.5 \mbox{ a.u.}=4.39 \mbox{ kHz } \mu\mbox{m}^3$. For a dielectric medium of refractive index $n$, we have an approximate expression 
$C_3=(n^2-1)/(n^2+1)C_3^{(\mathrm{metal})}$ \cite{Hoinkes}. 
For pure fused silica, 
with $n=1.45$ (in a broad region around the wavelength $\lambda_0=852$ nm), 
we find $C_{3g}=1.56 \mbox{ kHz } \mu\mbox{m}^3$. 
According to Ref. \cite{Ducloy},
the ratio between the van der Waals coefficients for excited- and ground-state cesium atoms is $C_{3e}/C_{3g}=1.98$. This yields $C_{3e}=3.09 \mbox{ kHz } \mu\mbox{m}^3$.
Regarding the potential depths, we use the experimental value 
$D_g=0.66 \mbox{ eV}=159.6 \mbox{ THz}$, measured as the adsorption energy of ground-state cesium atoms on fused silica \cite{Bouchiat}. 
With the assumption $D_e/D_g=C_e/C_g$,
we find $D_e=316 \mbox{ THz}$. For both the ground- and excited-state potentials, 
we find $x_a=0.21$ nm. Hence, the condition (\ref{29}) reads 
$0.13 \mbox{ nm}<x_m<0.21 \mbox{ nm}$. We assume that the minimum positions of  both potentials are the same and are equal to  $x_m=0.19$ nm. 
Then, Eq. (\ref{26}) yields $A_g=1.6\times 10^{18}$ Hz and $A_e=3.17\times 10^{18}$ Hz, while 
Eq. (\ref{28}) gives $\alpha_g=\alpha_e=53$ nm$^{-1}$.

\section{Density-matrix equations for a coherently driven atom}
\label{sec:driven}

We  present the
density-matrix equations for a coherently driven atom.
We assume that the atom is driven by one or several classical coherent plane-wave fields 
$\mathbf{E}_l$. Here the index $l$ labels the fields. For simplicity, we consider the case where all the driving fields propagate perpendicularly to the surface of the dielectric.
The expressions for the fields are given by 
$\mathbf{E}_l=[\mathcal{E}_l e^{i(\beta_lx-\omega_lt)}\boldsymbol{\epsilon}_l
+\mathrm{c.c.}]/2$,
where $\mathcal{E}_l$, $\boldsymbol{\epsilon}_l$, 
$\omega_l$, and $|\beta_l|=\omega_l/c$ are the envelopes, the polarization vectors, the frequencies, and the wave numbers, respectively. 
The quantities $\beta_l$ are positive or negative for rightward- or leftward-propagating fields, respectively.

The interaction between the atom and the driving fields is described by the Hamiltonian
\begin{equation}
H_I=-\frac{\hbar}{2}\sum_{lab}(\Omega_{lab}
e^{-i\omega_l t}|a\rangle\langle b|+\mathrm{H.c.}),
\label{4}
\end{equation}
where
\begin{equation}
\Omega_{lab}=\frac{\mathcal{E}_ld_{leg}}{\hbar} F_{ab}(\beta_l)
\label{5}
\end{equation}
is the Rabi frequency for the action of the field $\mathbf{E}_l$ on the transition 
between the translational levels $|a\rangle$ and $|b\rangle$. 
Here, $d_{leg}=\boldsymbol{\epsilon}_l\cdot\langle e|\mathbf{d}|g\rangle$ 
is the projection of the atomic dipole moment onto the polarization vector
$\boldsymbol{\epsilon}_l$, 
and $F_{ab}(k)=\langle \varphi_{a}|e^{ikx}| \varphi_{b}\rangle $
is the overlapping matrix element for the transition between the center-of-mass states $\varphi_{a}$ and $\varphi_{b}$ with a momentum transfer of $\hbar k$. 
It is interesting to note that, when we use the momentum representations
$\tilde{\varphi}_{a}(p)$ and $\tilde{\varphi}_{b}(p)$
for the wave functions ${\varphi}_{a}(x)$ and ${\varphi}_{b}(x)$, respectively, 
we have
\begin{equation}
F_{ab}(k)=\int_{-\infty}^{\infty}\tilde{\varphi}_{a}^*(p+\hbar k)\tilde{\varphi}_{b}(p) dp. 
\label{7}
\end{equation}
Equation (\ref{7}) indicates that the overlapping factor $F_{ab}(k)$ takes into account the 
change in the momentum of the atom.

When we apply the Schr\"{o}dinger equation $i\hbar\dot{\rho}=[H_I,\rho]$ to 
the Hamiltonian (\ref{4}) and include the radiative decay terms, 
described by Eqs. (\ref{66a}), we obtain the following equations for the density matrix $\rho$ of the atom:
\begin{eqnarray}
\dot{\rho}_{aa'}&=&\frac{i}{2}\sum_{l,b}(\Omega_{lab}\rho_{a'b}^* e^{-i\delta_{lab}t}
-\Omega_{la'b}^*\rho_{ab}e^{i\delta_{la'b}t})
\nonumber\\&&\mbox{} 
-\frac{1}{2}\sum_{a''}(\gamma_{a''a}e^{i\omega_{aa''}t}\rho_{a''a'}
+\gamma_{a''a'}^*e^{i\omega_{a''a'}t}\rho_{aa''}),
\nonumber\\
\dot{\rho}_{ab}&=&
\frac{i}{2}\sum_{l,b'}\Omega_{lab'}\rho_{b'b}e^{-i\delta_{lab'}t}
-\frac{i}{2}\sum_{l,a'}\Omega_{la'b}\rho_{aa'}e^{-i\delta_{la'b}t}
\nonumber\\&&\mbox{} 
-\frac{1}{2}\sum_{a'}\gamma_{a'a}e^{i\omega_{aa'}t}\rho_{a'b},
\nonumber\\
\dot{\rho}_{bb'}&=&-\frac{i}{2}\sum_{l,a}(\Omega_{lab'}\rho_{ab}^*e^{-i\delta_{lab'}t}
-\Omega_{lab}^*\rho_{ab'}e^{i\delta_{lab}t})
\nonumber\\&&\mbox{}
+\frac{1}{2}\sum_{aa'}(\gamma_{aa'bb'}+\gamma_{a'ab'b}^*)
e^{i(\omega_{bb'}-\omega_{aa'})t}{\rho}_{aa'}.
\label{23}
\end{eqnarray}
Here, $\delta_{lab}=\omega_l-\omega_a+\omega_b$ is the detuning of 
the driving-field frequency $\omega_l$ from the atomic transition frequency 
$\omega_{ab}$.

We note that Eqs. (\ref{23}) do not include the interaction of the atom with the phonons of the solid. We can take into account the decay due to phonons by adding
phenomenological terms \cite{Boyd}. The coefficients of such terms
are phonon absorption and emission probabilities and can be calculated in second-order perturbation theory \cite{Gortel}.

\end{document}